\documentstyle[manuscript,prl,aps]{revtex}
\begin{document}
\title{Equivalence of the Sutherland Model to Free Particles on a Circle} 
\author{N. Gurappa\thanks{panisprs@uohyd.ernet.in} and Prasanta K.
Panigrahi\thanks{panisp@uohyd.ernet.in}}  
\address{School of Physics, 
University of Hyderabad,
Hyderabad,\\ Andhra Pradesh,
500 046 INDIA.}
\maketitle

\begin{abstract} 
A method is developed to construct the solutions of one and many variable,
linear differential equations of arbitrary order. Using this, the
$N$-particle Sutherland model, with pair-wise inverse sine-square
interactions among the particles, is shown to be equivalent to free
particles on a circle. Applicability of our method to many other few and
many-body problems is also illustrated.

\end{abstract}
\draft
\pacs{PACS: 03.65.-w, 03.65.Ge}

\newpage
The Calogero-Sutherland model (CSM) \cite{cal,sut0} and its
generalizations \cite{sut,amp} are extremely well-studied quantum
mechanical systems, having relevance to diverse branches of physics
\cite{al}, such as the universal conductance fluctuations in mesoscopic
systems \cite{ms}, quantum Hall effect \cite{qhe}, wave propagation in
stratified fields \cite{wp}, random matrix theory \cite{sut0,al,naka,rmt},
fractional statistics \cite{fs}, gravity \cite{2dg} and gauge theories
\cite{gt}. In particular, the Sutherland model \cite{sut}, describing
$N$-identical particles having pair-wise inverse distance square
interaction on a circle has received wider attention in the context of
questions related to statistics \cite{poly}. The spectrum of this exactly
solvable model can be interpreted as arising from a set of free
quasi-particles satisfying the generalized exclusion principle
\cite{hald}.

Some time ago, the present authors have shown that, the CSM is equivalent
to a set of free harmonic oscillators \cite{prb}.  This mapping throws
considerable light into the algebraic structure of this correlated system
\cite{uji1,uji2}. Later, the same model, without the harmonic confinement,
was made unitarily equivalent to free particle \cite{gon}. However, the
connection between the Sutherland model and free particles have so far
remained an open problem. In this paper, we establish the equivalence of
the Sutherland model to free particles on a circle, for arbitrary number of
particles.

This paper is organized as follows. (i) A general method is developed to
solve any linear differential equation of arbitrary order, (ii) a few
examples are discussed to show how the present technique yields the known,
as well as, new results, (iii) the usefulness of this technique for
solving the Schr\"odinger equation with complicated potentials is pointed
out, (iv) the mapping between the $A_{N-1}$ Calogero model and free
harmonic oscillators is rederived using the present method, and, finally,
(v) we show, the equivalence of the Sutherland model with free particles
on a circle.

Consider the following differential equation, 
\begin{eqnarray} \label{ie}
\left(F(D) + P(x,d/dx)\right) y(x) = 0 \quad,
\end{eqnarray}
where, $D \equiv x \frac{d}{dx}$, $F(D) = \sum_{n = - \infty}^{n =
\infty} a_n D^n $, is a diagonal operator in the space of monomials
spanned by $x^n$ and $a_n$'s are some
parameters. $P(x,d/dx)$ can be an arbitrary polynomial function of $x$
and $\frac{d}{dx}$. We prove the following ansatz, 
\begin{eqnarray} \label{an}
y(x) &=& C_\lambda \left \{\sum_{m = 0}^{\infty} (-1)^m
\left[\frac{1}{F(D)}P(x,d/dx)\right]^m \right \} x^\lambda \nonumber\\
&\equiv& C_\lambda \hat{G}_\lambda x^\lambda \qquad,
\end{eqnarray}
is a solution of the above equation, provided, $F(D) x^\lambda = 0$ and
the coefficient of $x^\lambda$ in $y(x) - C_\lambda x^\lambda$ is zero
(no summation over $\lambda$);
here, $C_\lambda$ is a constant. The
case, when the equation $F(D) x^\lambda = 0$ does not have distinct roots
is not considered here and will be treated separately elsewhere.

Substituting Eq. (\ref{an}), modulo $C_\lambda$, in Eq. (\ref{ie}),
\begin{eqnarray}
\left(F(D) + P(x,d/dx)\right)&& \left\{\sum_{m = 0}^{\infty}
(-1)^m 
\left[\frac{1}{F(D)}P(x,d/dx)\right]^m \right \} x^\lambda = \nonumber\\
= F(D)&& \left[1 + \frac{1}{F(D)}P(x,d/dx)\right] \left \{\sum_{m =
0}^{\infty} (-1)^m \left[\frac{1}{F(D)}P(x,d/dx)\right]^m \right \}
x^\lambda \nonumber  \\
= F(D)&& \sum_{m = 0}^{\infty} (-1)^m
\left[\frac{1}{F(D)}P(x,d/dx)\right]^m  x^\lambda \nonumber\\
&& + F(D) \sum_{m = 0}^{\infty}(-1)^m \left[\frac{1}{F(D)}P(x,d/dx)
\right ]^{m + 1} x^\lambda \nonumber \\
= F(D)&& x^\lambda - F(D) \sum_{m = 0}^{\infty} (-1)^m
\left[\frac{1}{F(D)}P(x,d/dx)\right]^{m + 1} x^\lambda \nonumber\\
&& + F(D) \sum_{m = 0}^{\infty}(-1)^m \left[\frac{1}{F(D)}P(x,d/dx)
\right ]^{m + 1} x^\lambda \nonumber \\
= 0 \qquad.
\end{eqnarray}

Eq. (\ref{an}), which connects the solution of a given differential
equation to the monomials, can also be generalized to many-variables. In
order to show that, this rather straightforward procedure indeed yields
non-trivial results, we explicitly work out a few examples below and then
proceed to prove the equivalence of the Sutherland model to free particles
on a circle.

Consider the Hermite differential equation, which arises in the context
of quantum harmonic oscillator,
\begin{eqnarray}
\left(D - n - \frac{1}{2} \frac{d^2}{d x^2} \right) H_n(x) = 0 \qquad.
\end{eqnarray}
Here, $F(D) = D - n$ and $F(D) x^\lambda = 0$ yields $\lambda = n$.
Hence, 
\begin{eqnarray} 
H_n(x) = C_n \sum_{m = 0}^{\infty}(-1)^m \left[\frac{1}{D - n} (-
1/2) (d^2/dx^2) \right ]^{m} x^n \qquad.
\end{eqnarray}
Using, $[D \,\,,\,\,(d^2/dx^2)] = -2 (d^2/dx^2)$, it is easy to see that,
\begin{eqnarray}
\left[\frac{1}{(D - n)} (- 1/2) (d^2/dx^2) \right ]^{m} x^n = (- 1/2)^m
(d^2/dx^2)^m \prod_{l=1}^m \frac{1}{(- 2 l)} x^n \qquad, \nonumber
\end{eqnarray}
and 
\begin{eqnarray}
H_n(x) &=& C_n \sum_{m = 0}^{\infty}(-1/4)^m \frac{1}{m!} (d^2/dx^2)^m
x^n \nonumber\\  
&=& C_n e^{- \frac{1}{4} \frac{d^2}{dx^2}} x^n \qquad;
\end{eqnarray}
this is a well-known result. Similar expression also holds for the Lagurre
polynomials which matches with the one found in \cite{ak}. In order to
make an important remark, we list the solutions of some frequently
encountered differential equations in various branches of physics
\cite{gra}.

\noindent \underline{Legendre polynomials}
\begin{eqnarray}
P_n(x) = C_n e^{- \left\{1/(2[D + n + 1]) \right\} (d^2/d x^2)}
\,\, x^n \qquad.\nonumber
\end{eqnarray}

\noindent \underline{Associated Legendre polynomials}
\begin{eqnarray}
P_n^m (x) = C_n (1 - x^2)^{m/2} e^{- \left \{1/(2[D + n + m +
1])\right\}(d^2/d x^2)} \,\, x^{n - m} \qquad.\nonumber
\end{eqnarray}


\noindent \underline{Bessel functions}
\begin{eqnarray}
J_{\pm \nu}(x) = C_{\pm \nu} e^{- \{1/(2[D \pm \nu])\}x^2} \,\,x^{\pm
\nu} \qquad.\nonumber 
\end{eqnarray}

\noindent \underline{Generalized Bessel functions}
\begin{eqnarray}
u_{\pm}(x) = C_{\pm} e^{-\{ \beta \gamma^2 /(2 [D + \alpha \pm
\beta \nu ])  \}x^{2 \beta}} \,\, x^{\beta \nu - \alpha \pm \beta \nu}
\qquad.\nonumber 
\end{eqnarray}

\noindent \underline{Gegenbauer polynomials}
\begin{eqnarray}
C_n^\lambda(x) = C_n e^{- \{1/(2[D + n + 2 \lambda ])\} (d^2/d x^2)}
\,\,x^n \qquad.\nonumber
\end{eqnarray}

\noindent \underline{Hypergeometric functions}
\begin{eqnarray}
y_{\pm}(\alpha, \beta ; \gamma; x) = C_{\pm} e^{- \{(1/ (D +
\lambda_{\pm}) \}\hat{A}} \,\,x^{- \lambda_{\mp}} \qquad,\nonumber
\end{eqnarray}
where, $\lambda_{\pm}$ is either $\alpha$ or $\beta$ and $\hat{A}
\equiv x \frac{d^2}{d x^2} + \gamma \frac{d}{d x}$. All the above series 
solutions have descending powers of $x$. In order to get the series in the
ascending powers, one has to replace $x$ by $\frac{1}{x}$ in the original
differential equation and generate the solutions via Eq. (\ref{an}).
However, the number of solutions will remain the same. One can also
generate the series solutions by multiplying the original differential
equations with $x^2$, and then, rewriting $x^2 \frac{d^2}{d x^2} = D (D
- 1) = F(D)$.

\noindent \underline{Meijer's G-Function} 
\begin{eqnarray}
G_{pq}^{mn}\left(x \mid_{b_s}^{a_r}\right) = \left\{\sum_{k = 0}^\infty 
\left[\frac{1} {\prod_{j=1}^q \left(D - b_j \right)} \left((-1)^{p - m
- 1} x \prod_{j=1}^p \left( D - a_j + 1 \right)\right) \right]^k
\right\}\,\, x^{b_s} 
\qquad.\nonumber
\end{eqnarray}

\noindent \underline{Neumann's polynomials}
\begin{eqnarray}
O_n(x) = \left\{\sum_{r=0}^{\infty} (- 1)^r \left[ \frac{1}{[(D + 1)^2
- n^2]} x^2 \right]^r \left(\frac{1}{[(D + 1)^2 - n^2]} \right) \right\}
\,\, (x \cos^2 (n \pi/2) \nonumber\\
\qquad \qquad \qquad \qquad \qquad \qquad \qquad \qquad \qquad  + n
\sin^2(n \pi/2) ) \,\,; \nonumber  
\end{eqnarray}
the cases of Struve, Lomel, Anger and Weber functions are identical to the
above one. Jacobi, Schl\"afli, Whittaker, Chebyshev and some other
polynomials were not given here since the list is rather lengthy.  Now,
the remark follows: $\hat{G}_\lambda$ becomes independent of the roots of
the equation $F(D) x^\lambda = 0$, only when $F(D)$ is linear in $D$.

The solution for the following equation with periodic potential,
\begin{eqnarray} \label{pp}
\frac{d^2 y}{d x^2} + a \cos(x) y = 0 \quad,
\end{eqnarray}
can be found, after multiplying Eq. (\ref{pp}) by $x^2$ and rewriting $x^2 
\frac{d^2}{d x^2}$ as $(D - 1) D$, to be
\begin{eqnarray}
y(x) = \sum_{m , \{n_i\} = 0}^{\infty} \frac{(- a)^m}{m!}
\left\{\prod_{i = 1}^m \frac{(-1)^{n_i}}{(2n_i)!} \right\}&& 
\left\{\prod_{r = 1}^m \frac{(2 [m + \lambda/2 - r + 
\sum_{i=1}^{m+1-r}n_i])!} {(2 [m + \lambda/2 + 1 - r +
\sum_{i=1}^{m+1-r}n_i])!} 
\right\} \nonumber\\ 
&& \times \,\,x^{2(m + \sum_{i=1}^m n_i + \lambda/2)} \qquad,
\end{eqnarray}
where, $\lambda = 0$ or $1$. In the same manner, one can write down the
solutions for the Mathieu's equation as well \cite{gra}.

Using this method, one can, in principle write down the ground and first
excited states of the one variable Schr\"odinger equation. We first
illustrate the procedure for harmonic oscillator, ($\hbar = m = \omega =
1$),
\begin{eqnarray}
\left(\frac{d^2}{d x^2} + (2 E_n - x^2)\right) \psi_n = 0 \qquad,
\end{eqnarray}
before proceeding to the other non-trivial examples. Multiplying by
$x^2$, the above can be written as
\begin{eqnarray}
\left((D - 1) D + x^2 (2 E_n - x^2)\right) \psi_n = 0 \qquad.
\end{eqnarray}
For $n = 0$ and $1$, $D (D - 1) x^n = 0$. Using Eq. (\ref{an}), the
solution for $n = 0$ is,
\begin{eqnarray}
\psi_0 &=& C_0 \left \{\sum_{m = 0}^{\infty} (-1)^m
\left[\frac{1}{(D - 1) D} (x^2 (2 E_0 - x^2))\right]^m \right \}
x^0 \nonumber\\
&=& C_0 \left(1 - \frac{[2 E_0]}{2 !} x^2 + \frac{(2! + [2 E_0]^2)}{4!}
x^4 - \frac{(4! + (2!)^2 [2 E_0] + (2!) [2 E_0]^3)}{2! 6!} x^6 + \cdots
\right) \,\,.
\end{eqnarray}
Note that, $\psi_0$ is an expansion in the powers of $x$, whose
coefficients are polynomials in $E_0$. Only when $E_0 = 1/2$, the
series can be written in the closed form $C_0 e^{- \frac{1}{2} x^2}$,
which satisfies all properties required for a quantum mechanical
ground-state.  Similarly, the first excited state can also be found by
applying $\hat{G}_1$ on $x$. To find the $n$-th excited state, one can
differentiate the Schr\"odinger equation $n$ times, multiply it by
$x^n$ and use $x^n \frac{d^n}{d x^n} =
\prod_{l = 0}^{n - 1} (D - l) = F(D)$.  In the case of complicated
potentials, this method can be potentially very useful if applied in
conjunction with numerical algorithms. The exact ground-state wavefunction
for the Schr\"odinger equation with anharmonic potential, $V(x) = x^2/2 +
c x^4/2$, is,
\begin{eqnarray}
\psi_0 = C_0 \left \{\sum_{m = 0}^{\infty} (-1)^m
\left[\frac{1}{(D - 1) D} (x^2 (2 E_0 - x^2 - c x^4))\right]^m
\right \}\,\,\,. 1 \qquad.
\end{eqnarray}
One can numerically tune $E_0$ to obtain an appropriate $\psi_0$.

In the following, we apply the above technique to the differential
equations involving many variables. Consider, the $A_{N-1}$
Calogero-Sutherland model, which was made equivalent to free harmonic
oscillators \cite{prb},
\begin{eqnarray}
\left(- \frac{1}{2} \sum_{i=1}^N \frac{{\partial}^2}{\partial x_i^2}
+ \frac {1}{2} \sum_{i=1}^N x_i^2 + \frac {1}{2} g^2 \sum_{{i,j=1}\atop
{i\ne j}}^N \frac {1}{(x_i - x_j)^2} - E_n\right) (\psi_0 \phi_n) = 0
\qquad. 
\end{eqnarray}
This can be brought to the following form, after the removal of the
ground-state wavefunction, $\psi_0 = \exp\{-\frac{1}{2} \sum_i x_i^2\}
\prod_{i<j}^N [|x_i - x_j|^\alpha $:
\begin{eqnarray} \label{til}
\left(\sum_i x_i \frac{\partial} {\partial x_i} +  E_0 - E_n - \hat A
\right) \phi_n = 0\qquad,  
\end{eqnarray}
where, $E_0 = \frac{1}{2} N + \frac{1}{2} N (N-1) \alpha $ is the
ground-state energy and $\hat A \equiv [\frac{1}{2}
\sum_i\frac{\partial^2}{\partial x_i^2} + \alpha \sum_{i\ne j}
\frac{1}{(x_i - x_j)} \frac{\partial}{\partial x_i}]$. Without loss of
generality, we have quantized the above system as bosons.  Rewriting Eq.
(\ref{til}) as,
\begin{eqnarray}
\left(\sum_i x_i \frac{\partial}{\partial x_i} - n + n + E_0 - E_n -
\hat A \right) \phi_n = 0\qquad, 
\end{eqnarray}
the solution can be written by using Eq. (\ref{an}):
\begin{eqnarray}
\phi_n = C_n \left\{\sum_{m = 0}^{\infty} (-1)^m
\left[\frac{1}{(\sum_i x_i \frac{\partial}{\partial x_i} - n)} 
(n + E_0 - E_n - \hat{A}) \right]^m \right\} S_n(\{x_i\}) \quad;
\end{eqnarray}
here, $(\sum_i x_i \frac{\partial}{\partial x_i} - n) S_n(\{x_i\}) = 0$
and $S_n ( \{x_i \})$'s are any homogeneous function of degree $n$. In
order to avoid the possibility of having singular solutions when the
inverse of $(\sum_i x_i \frac{\partial}{\partial x_i} - n)$ acts on
$S_n(\{x_i \})$, we choose $n + E_0 - E_n = 0$; this yields the familiar
energy spectrum of the Calogero-Sutherland model.  Further, one has to
choose $S_n(\{x_i \})$ to be completely symmetric under the exchange of
$x_i$'s, such that, the action of $\hat A$ yields polynomial solutions
which are normalizable with respect to $\psi_0$ as the weight
function\cite{prb,bak}. Similar to the case of the Hermite polynomial, one
can easily prove that,
\begin{eqnarray}
\phi_n = C_n e^{- \frac{1}{2} \hat A } S_n(\{x_i \}) \qquad.
\end{eqnarray}

In the following, we show that, the Sutherland model is equivalent to
free particles on a circle. The Schr\"odinger equation is, 
\begin{eqnarray} \label{sut1}
\left(- \sum_{i=1}^N \frac{\partial^2}{ \partial x_i^2} +
2 \beta (\beta - 1) \frac{\pi^2}{L^2} \sum_{i<j} \frac{1}{\sin^2[\pi (x_i
- x_j)/L]} - E_\lambda \right) \psi_\lambda(\{x_i\}) = 0 \qquad.
\end{eqnarray}
Choosing, $z_j = e^{2\pi i x_j /L}$ and writing $\psi_\lambda(\{z_i\})
= \prod_i z_i^{- (N - 1) \beta /2} \prod_{i<j} (z_i - z_j)^\beta
J_\lambda(\{z_i\})$, the above equation becomes,
\begin{eqnarray} \label{jac}
\left(\sum_i D_i^2 + \beta \sum_{i<j} \frac{z_i + z_j}{z_i - z_j}
(D_i - D_j) + \tilde{E}_0 - \tilde{E}_\lambda \right)
J_\lambda(\{z_i\}) = 0 \qquad, 
\end{eqnarray}
where, $D_i \equiv z_i \frac{\partial}{\partial z_i}$, 
$\tilde{E_\lambda} \equiv (\frac{L}{2 \pi})^2 E_\lambda$,
$\tilde{E_0} \equiv (\frac{L}{2 \pi})^2 E_0$ and $E_0 =
\frac{1}{3}(\frac{\pi}{L})^2 \beta^2 N (N^2 - 1)$, is the ground-state
energy. Here, $J_\lambda(\{z_i\})$ is known
as the Jack polynomials \cite{jack1,jack2,jack3,scha}.
$\sum_i D_i^2$ is a diagonal operator in the space spanned by the
monomial symmetric functions, $m_{\{\lambda\}}$, with eigenvalues
$\sum_{i=1}^N \lambda_i^2$ \cite{jack3}. Rewriting (\ref{jac}) in the
form, 
\begin{eqnarray} 
\left(\sum_i (D_i^2 - \lambda_i^2) + \beta \sum_{i<j} \frac{z_i +
z_j}{z_i - z_j} (D_i - D_j ) + \tilde{E}_0 + \sum_i \lambda_i^2 -
\tilde{E}_\lambda \right) J_\lambda(\{z_i\}) = 0 \qquad, 
\end{eqnarray}
one can immediately show that,
\begin{eqnarray} \label{sr}
J_\lambda(\{z_i \}) &=& C_\lambda \left \{\sum_{n = 0}^{\infty} (-1)^n
\left[\frac{1}{\sum_i (D_i^2 - \lambda_i^2)}(\beta \sum_{i<j} \frac{z_i +
z_j}{z_i - z_j}(D_i - D_j ) +  \tilde{E}_0 + \sum_i \lambda_i^2 -
\tilde{E}_\lambda)\right]^n \right \} \nonumber\\ 
&& \qquad \qquad \qquad \qquad \qquad \qquad \qquad \qquad \qquad
\qquad \qquad \qquad \qquad \times  m_\lambda(\{z_i\}) \nonumber\\ 
&\equiv& C_\lambda \hat{G}_\lambda m_\lambda(\{z_i \}) \qquad.
\end{eqnarray}
It is easy to check that, $\hat{G}_\lambda$ maps
the Sutherland model to free particles on a circle, {\it i.e.},
\begin{eqnarray}
(\psi_0 \hat{G}_\lambda)^{-1} H_S (\psi_0 \hat{G}_\lambda) &=&
\left(\frac{2 \pi}{L}\right)^2 \left(\sum_i D_i^2 - \sum_i \lambda_i^2 +
\tilde{E}_\lambda \right) \nonumber\\
&=& - \sum_{i=1}^N
\frac{\partial^2}{ \partial x_i^2} - \left(\frac{2 \pi}{L}\right)^2
\sum_i \lambda_i^2 + {E}_\lambda \qquad, 
\end{eqnarray}
where, $H_S$ is the Sutherland Hamiltonian and $\psi_0$ is its
ground-state wavefunction. For the sake of convenience, we define
\begin{eqnarray} \label{S}
\hat{S} & \equiv & \left[\frac{1}{\sum_i (D_i^2 - \lambda_i^2)}
\hat{Z} \right] \qquad, \nonumber\\
\mbox{and} \qquad
\hat{Z} & \equiv & \beta \sum_{i<j} \frac{z_i + z_j}{z_i - z_j}
(D_i - D_j) +  \tilde{E}_0 + \sum_i \lambda_i^2 - \tilde{E}_\lambda
\qquad.  
\end{eqnarray}
The action of $\hat{S}$ on $m_\lambda(\{z_i \})$ yields singularities,
unless one chooses the coefficient of $m_\lambda$ in $\hat{Z} \,
m_\lambda(\{z_i\})$ to be zero; this condition yields the well-known 
eigenspectrum of the Sutherland model:
$$
\tilde{E}_\lambda = \tilde{E}_0 + \sum_i (\lambda_i^2 + \beta [N + 1 -
2 i] \lambda_i) \qquad. 
$$ 
Using the above, one can write down the Jack polynomials as,
\begin{eqnarray} \label{nfj}
J_\lambda(\{z_i \}) = \sum_{n=0}^\infty (- \beta)^n 
\left[\frac{1}{\sum_i (D_i^2 - \lambda_i^2)} (\sum_{i<j} \frac{z_i +
z_j}{z_i - z_j}(D_i - D_j) - \sum_i  
(N + 1 - 2 i) \lambda_i )\right]^n \nonumber\\
\qquad \qquad \qquad \qquad \qquad  \times m_\lambda(\{x_i\}) \qquad.
\end{eqnarray}
For the sake of illustration, we give the following 
computation of $J_{2}$ for two-particle case.
Now, $m_{2} = z_1^2 + z_2^2$ and it is easy to check that,
\begin{eqnarray}
\hat{Z} m_{2} &=& \beta 4 z_1 z_2 = 4 \beta m_{1^2} \nonumber\\
\hat{S} m_{2} &=& \frac{1}{\sum_i (D_i^2 - 4)} (4 \beta m_{1^2}) = - 2
\beta m_{1^2} \qquad, \nonumber\\
\hat{S}^n  m_{2} &=& - 2 (\beta)^n m_{1^2} \quad \mbox{for}\quad n \ge 1
\quad. \nonumber 
\end{eqnarray}
Substituting the above result in Eq. (\ref{sr}), apart from $C_2$,
\begin{eqnarray}
J_{2} &=& m_{2} + \left(\sum_{n=1}^\infty (- 1)^n (-2) (\beta)^n \right)
m_{1^2} \nonumber\\
&=& m_{2} + 2 \beta \left(\sum_{n = 0}^\infty (- \beta)^n \right)
m_{1^2} \nonumber\\
&=& m_{2} + \frac{2 \beta}{1 + \beta} m_{1^2} \qquad;
\end{eqnarray}
this is the desired result. Earlier, Lapointe {\it et al.} have obtained a
Rodrigues-type formulae for the Jack polynomials \cite{lllv}. Eq.
(\ref{nfj}) is a new formula for the Jack polynomials. In the
following, we show that, the above method can be generalized to more
general many-body systems.

Consider,
\begin{eqnarray}
\left(\sum_{n = -\infty}^\infty a_n (\sum_i D_i^n) + \hat{A} \right)
P_\lambda(\{z_i\}) = B_\lambda(\{z_i\}) \qquad,
\end{eqnarray}
where, $a_n$'s are some parameters, $D_i$ is same as the above, $\hat{A}$
is a function of $z_i$ and $\partial/\partial z_i$ and
$B_\lambda(\{z_i\})$ is a source term. 

\noindent {\it Case (i)} When $B_\lambda(\{z_i\})= 0$ and $\hat{A}
m_\lambda = 
\epsilon_\lambda m_\lambda + \sum_{\mu < \lambda} C_{\mu \lambda}
m_{\mu}$; where, $m_\lambda$'s are the monomial symmetric functions
\cite{jack3} and $\epsilon_\lambda$ and $C_{\lambda \mu}$ are some
constants, we get
\begin{eqnarray}
P_\lambda(\{z_i\}) = \sum_{r=0}^\infty (- 1)^r \left[\frac{1}{((\sum_{n =
-\infty}^\infty a_n (\sum_i D_i^n) - (\sum_{n = -\infty}^\infty a_n
(\sum_i {\lambda}_i^n))} (\hat{A} - \epsilon_\lambda)\right]^r
m_\lambda(\{z_i\}) \,\,
\end{eqnarray}
with, $\sum_{n = -\infty}^\infty a_n (\sum_i {\lambda}_i^n) + 
\epsilon_ \lambda = 0$.

\noindent {\it Case (ii)} When $B_\lambda(\{z_i\})= \sum_{\mu \ne \lambda}
K_{\mu} m_\mu$ and $\hat{A} m_\mu = \sum_{\mu \ne \lambda} C_{\mu \lambda}
m_{\mu}$; where, $K_\mu$ and $C_{\lambda \mu}$ are some
constants, then,
\begin{eqnarray}
P_\lambda(\{z_i\}) = \sum_{r=0}^\infty (- 1)^r \left[\frac{1}{((\sum_{n =
-\infty}^\infty a_n (\sum_i D_i^n) - (\sum_{n = -\infty}^\infty a_n
(\sum_i {\lambda}_i^n))} \hat{A}\right]^r \nonumber\\
\times \left[\frac{1}{((\sum_{n =
-\infty}^\infty a_n (\sum_i D_i^n) - (\sum_{n = -\infty}^\infty a_n
(\sum_i {\lambda}_i^n))}\right] B_\lambda(\{z_i\}) \,\,.
\end{eqnarray}

In conclusion, we have developed a general method to solve any linear
differential equation of arbitrary order. The major advantage of this
technique lies in the fact that, in order to obtain the solutions, one
does not have to solve any recursion relations or integral equations. By
applying this method, we obtained new formulae for many differential
equations which are of interest to theoretical physicists. We also
discussed how to obtain the ground and excited states for the
Schr\"odinger equations with complicated potentials. In particular, we
have written down the exact ground-state wavefunction for anharmonic
potential, $V(x) = x^2/2 + c x^4/2$, as a function of energy parameter,
which has to be tuned to obtain the normalizable solutions. This method
treated both $A_{N-1}$ Calogero-Sutherland and the Sutherland model with
inverse sine-square interactions on the equal footing. The former one is
equivalent to free harmonic oscillators and the later, to free particles
on a circle. We also proposed a general scheme for generating more general
symmetric functions. This scheme may find interesting applications in the
context of obtaining the $q$-deformed versions of the known polynomials
\cite{gra,jack3}.

The authors acknowledge extremely useful discussions with Profs. V. 
Srinivasan and S. Chaturvedi. N.G thanks U.G.C (India) for the
financial support.

\end{document}